\documentclass[twocolumn,showpacs,amsmath,amssymb,aps]{revtex4}
\usepackage{graphicx,amsmath}
\usepackage{txfonts}

\begin{document}
\title{Cooperation, structure and hierarchy in multiadaptive games}

\author{Sungmin Lee}
\affiliation{IceLab, Department of Physics, Ume{\aa} University, 90187 Ume\aa, Sweden}

\author{Petter Holme}
\affiliation{IceLab, Department of Physics, Ume{\aa} University, 90187 Ume\aa, Sweden}
\affiliation{Department of Energy Science, Sungkyunkwan University, Suwon 440--746, Korea}
\affiliation{Department of Sociology, Stockholm University, 10691 Stockholm, Sweden}

\author{Zhi-Xi Wu}
\affiliation{Institute of Computational Physics and Complex Systems, Lanzhou University, Lanzhou, Gansu 730000, China}

\begin{abstract}
Game-theoretical models where the rules of the game and the interaction structure both coevolves with the game dynamics---multiadaptive games---capture very flexible situations where cooperation among selfish agents can emerge. In this work, we will discuss a multiadaptive model presented in a recent Letter [Phys.\ Rev.\ Lett.\ \textbf{106}, 028702 (2011)], and generalizations of it. The model captures a non-equilibrium situation where social unrest increases the incentive to cooperate and, simultaneously, agents are partly free to influence with whom they interact. First, we investigate the details of how the feedback from the behavior of agents determines the emergence of cooperation and hierarchical contact structures.  We also study the stability of the system to different types of noise, and find that different regions of parameter space show very different response. Some types of noise can destroy an all-cooperator state. If, on the other hand, hubs are stable, then so is the all-C state. Finally, we investigate the dependence of the ratio between the timescales of  strategy updates and the evolution of the interaction structure. We find that a comparatively fast strategy dynamics is a prerequisite for the emergence of cooperation.
\end{abstract}
\pacs{02.50.Le,89.75.Hc,89.75.Fb,87.23.Ge}
\maketitle

\section{Introduction}

An open question in both biology and the social sciences is how cooperation and the network of social interactions co-emerge in a population of selfish individuals. Game theory is the basic theoretical framework to investigate such phenomena~\cite{axe:evo_maynard_nowak:evodyn}. Furthermore, game theory is a language for describing systems in biology, economy and society where the success of an agent depends both on its own behavior and the behaviors of others. Through the development and the study of different types of models, researchers have captured different game-theoretic scenarios. Our previous work~\cite{ourPRL} showed a new direction, relaxing constraints of other models with feedback at different levels to the behavior of the agents. This class of systems can be anticipated to show a rich behavior and be interesting for interdisciplinary studies of social systems. In this paper, we study the multiadaptive game of Ref.~\cite{ourPRL} in greater detail and, primarily, extend and simplify it in various ways to paint a fuller picture of the class of models that the model in Ref.~\cite{ourPRL} belongs to.

We motivate our multiadaptive model as an extension of spatial social dilemmas---systems driven by a conflict between collective and individual interests, and the interaction happen between agents that are close in space. More specifically, we start from the Nowak--May (NM) game~\cite{nowakmay} that is technically on the border between the Prisoner's dilemma and Chicken games. It captures social situations where at any time it is most rewarding to defect. However, in some situations, in a long-term perspective,  agents benefit from establishing mutual cooperation. More mathematically, each agent can take two actions: defect (D) or cooperate (C). Cooperation means, in this context at least, that the agents do what is best for the community. An encounter in the NM game gives zero payoff to anyone interacting with a defector (D), payoff one to a cooperator (C) meeting another cooperator, and $b>1$ to a D meeting a C. In the literature, people have used this model to explain the emergence of cooperation among selfish agents in a vast number of disciplines---political science, economics, and biology~\cite{axe:evo_maynard_nowak:evodyn}.

In the original NM game, the game rules, as parameterized by the payoff matrix, are fixed in time.  In real-world  systems, there could well be a feedback mechanism from the overall success of the agents, i.e.\ the society, to the payoff matrix. Imagine for instance that there is a stable, widespread cooperation that builds up a common wealth among the agents. In such a situation, there would be more common resources at stake at every interaction, and thus a larger temptation to defect. The easiest way to incorporate feedback from the entire system to the game rules is to let the entries of the matrices be variables that are dependent on external environment (the society in socioeconomic game theory, the environment for evolutionary models). This is what we will do and, following Ref.~\cite{tomochi}, we define
\begin{equation}\label{tupdate}
b(t+1)=b(t)+\alpha[\rho (t)-\rho^*],
\end{equation}
where $t$ is the discrete simulation time, $\rho$ is the fraction of cooperators, $\rho^*\in[0,1]$ is a model parameter signifying a neutral cooperation level, and $\alpha>0$ sets the strength of the feedback from the environment to the payoff. The idea behind this form is that a high cooperation level means the society gets rich which should increase the incentive to exploit this richness, and thus increase $b$. This is not supposed to be regarded as a universal mechanism; rather a scenario that could apply to a restricted set of  social or environmental situations. The linear response form is motivated by simplicity; one could also imagine a threshold response (in analogy to other models of response to social influence~\cite{influence}).

The other feature we leave flexible and adaptive is the interaction structure (cf.\ Refs.~\cite{egui:evo_ptn:coevo_zsch,wli2007,zimmermann2000}), i.e.\ the social network between the agents. In the spirit of the ``strength of week ties'' idea~\cite{grano:weak}, we assume the environment of people can be differentiated into strong local ties to family and work colleagues that are hard to break and of little use when it comes to changing ones social situation, and weak ties that helps to reach information, or to build new social ties, further in the social network (the motivating example was how people find new jobs). In our model, we will also have local ties that do not change. To keep the similarity to the NM game we let them be the four neighbors of a square grid. In addition to the local ties, each agent has one connection that could reach anyone outside of the neighborhood. The agents, we assume, use this ``weak tie'' to optimize their position in the social network by connecting it to it best-performing neighbor (including neighbors of the weak, long-range link, so that this link can wander off, away from the local surrounding of strongly connected links). This setup, inspired by Ref.~\cite{wli2007} will be described more algorithmically below.

The rest of the paper is organized as follows. In Sec.~\ref{static}, we define what we call the adaptive model essentially the scenario above without the social-network dynamics. In this section, we also analyze this model numerically.  In section~\ref{mmodel} we present and investigate the full, multiadaptive model, including the emergence of cooperation and social structure.  In Sec.~\ref{wnoise}, we study the response of a system on noise and, particularly, the stability of all-C state. We investigate the role of timescale differences between updating strategy and rewiring non-local links in Sec.~\ref{tscale}. Finally, in Sec.~\ref{discussion}, we discuss our results and open problems.

\begin{figure}
\includegraphics[width=\linewidth]{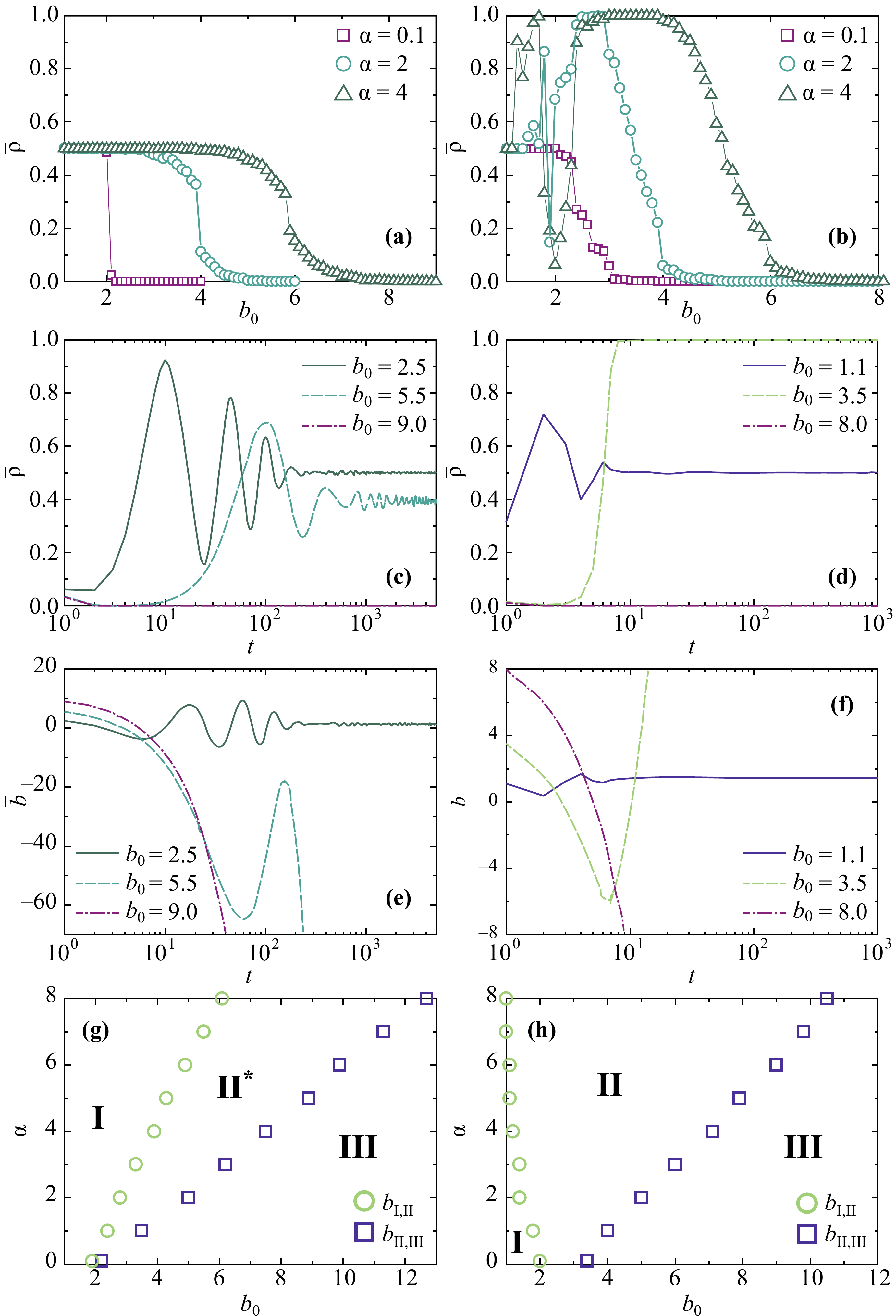}
\caption{(Color online) Parameter dependence of the game as seen in the temptation and average cooperator density on a square lattice (left panels). Right panels shows the results adding non-local links. Panels (a) and (b) show average density of cooperators $\bar\rho $, as a function of the initial temptation $b_0$, with $\alpha=0.1$, $2$, and $4$. The bar represents points averaged over the last $1000$ of $5000$ (c) and (e) and $500$ of $1000$ (d) and (f) steps. Panels (c) and (d), and (e) and (f)  correspond to the time evolution of $\bar\rho $ and $\bar b$, respectively, for  different values of $b_0$ when $\alpha=4$. Panels (g) and (h) show the diagram over the three regions in $\alpha$--$b_0$ space. The curves are averages over $10^4$ runs.}\label{onstatic}
\end{figure}

\section{Adaptive model}\label{static}

In this section, we will successively move from the NM game toward the multiadaptive game mentioned above. First, we will include feedback from the environment---the overall wealth of the agents---to the rules of the games as parameterized by the payoff matrix. Later we will investigate the case where an agent has a long-range link (which, in this section, is not open to optimization).

The basic set-up is an $L\times L$ square grid of agents interacting with their four nearest neighbors. We let this square grid  have  fixed boundary conditions, so that an agent in the interior interacts with four others, an agent on the links interact with three others and an agent in the corners interact with two others.  Unless otherwise stated, we will use $L=100$. As mentioned above, the global state of the system is determined by the temptation to defect $b(t,\rho)$ given by Eq.~(\ref{tupdate}). The initial value $b_0$ of the temptation is another parameter value of the model.

Starting from a random configuration of defectors and cooperators in the population, we update the system first by calculating the total payoff of an agent $i$ as the sum of the payoffs obtained in the last, synchronous, interaction,
\begin{equation}\label{payoff}
u_i=\sum^N_{j=1} u_{ij} A_{ij}
\end{equation}
where $u_{ij}$ is $i$'s payoff from the NM game in the interaction with $j$, and $A_{ij}$ is the adjacency matrix, whose value is $1$ (or $0$) depending on whether (or not) $i$ and $j$ are neighbors in the social network. When the payoff is gleaned the agents, once again synchronously, chose strategies (whether to cooperate or defect). If an agent $i$ has a higher total payoff than its neighbors nothing happens. But, if an agent $j$, with an link to $i$ has a higher total payoff, then $i$ use the same strategy in the next timestep as $j$ just did with a transition probability given by~\cite{wli2007},
\begin{equation}\label{rewiringprob}
\Pi(i\rightarrow j)=\frac{1}{1+e^{-\beta (u_j-u_i)}}~.
\end{equation}
Here $\beta$ parameterizes the noise in the choice of whom to imitate. This type of selection noise is further discussed in Ref.~\cite{szabotoke}. Except in Sec.~\ref{wnoise}, we use $\beta=1$.

In Fig.~\ref{onstatic}, we give numerical results summarizing the behavior of the model. In panel (a), we display the average cooperator density, $\bar\rho$, as a function of $b_0$ for three values of $\alpha$. If $\alpha=4$ and $b_0 < 4.0$, we can see the system converging to a state with $\bar\rho\approx\rho^*=1/2$. This behavior holds through a region of the parameter space that we denote I.  For large  $b_0$-values, there is a transfer to another behavior---region III---characterized by a vanishing cooperation. We call the value where this happens  $b_{\mathrm{I,III}}$. $b_{\mathrm{I,III}}$, we note, increase with $\alpha$. This means that if the coupling between the overall behavior and the payoff matrix gets stronger, defection needs higher initial temptation values to take over the population. Between regions I and III there is a region $\mathrm{II}^*$ where, depending on $b_0$, the cooperation density either converges to $\rho^*$ or $0$ with probabilities depending on $\alpha$ and $b_0$. (We save the notation $\mathrm{II}$ for another behavior that will be discussed below.)  With increasing $b_0$, the probability that the system ends in $\rho^*$ decreases, and vanishes completely at $b_{\mathrm{II^*,III}}$. In Figs.~\ref{onstatic}(c) and (e), we display trajectories of $\rho$ and $b$, averaged over $10^4$ runs, for $b_0=2.5$, 5.5, and 9.0 with $\alpha=4$. These curves show that the system stabilizes to a steady cooperation level after an oscillatory transient. In terms of configurations, such  oscillations are manifested as growing and shrinking C or D clusters. This can be explored further with our Java applet of the model~\cite{java}. For all parameter values we study, the oscillatory behavior will be dampened to a fixed point at (or, at least, very close to) $\rho^*$. An interesting observation is that in region $\mathrm{I}$ the temptation can be  controlled by the feedback so that the final density of cooperators $\rho^*$ is at some intermediate value between $1$ and  $0$ (that is, both C and D clusters coexist in the fixpoint). Dynamically,  region $\mathrm{III}$ is characterized by the system hitting the fixed point $\rho=0$ faster than the environment can respond by tuning the value of $b$. In Fig.~\ref{onstatic}(g), we plot the boundaries between the regions in the $\alpha$--$b_0$ plane.  We  identify region I numerically as when $\bar\rho$, at convergence, deviates less than $0.5\%$ from $\rho^*$, in other words that $|\bar\rho-\rho^*|$ is less than half a percent. The  region $\mathrm{III}$ identified to when the  $\bar \rho < 0.005$. From Fig.~\ref{onstatic}(g) we see that  the boundary value, $b_{\mathrm{I,II^*}}$, separating region $\mathrm{I}$ from $\mathrm{II}^*$ increases with an increasing $\alpha$---the coexistence region for $C$ and $D$ becomes wider with increasing  strength of the feedback. In region I, for all measured values of $\alpha$, the system relaxes to a steady state with $\bar\rho\approx\rho^*$ and $\bar b$ converges to an intermediate  value. For example, $b_0=2.5$ gives $b(t\rightarrow\infty)\simeq 1.2$ [Figs.~\ref{onstatic}(c) and (e)]. This happens when the feedback in Eq.~(\ref{tupdate}) is strong enough to balance $b$. When $b_0$ increases beyond $b_{\mathrm{I,II^*}}$, the feedback from the environment starts affecting $b$ so much that the system  hits the all-D state. As a final note on Fig.~\ref{onstatic}(g), we see that $b_{\mathrm{II^*,III}}$, separating region $\mathrm{II^*}$ from $\mathrm{III}$, increases monotonically with $\alpha$.

Next, we continue moving closer to the multiadaptive game by introducing long-range links to every agent. At this stage, they are distributed randomly and not open to optimization. When the strength of feedback is weak enough ($\alpha=0.1$), $\bar\rho$ shows a  behavior similar to the case without non-local links [see Fig.~\ref{onstatic}(b)]. However, the average density of cooperators $\bar\rho$ as a function of $b_0$  changes drastically when $\alpha$ is larger ($\alpha=2$ and $4$ in Fig.~\ref{onstatic}). For example,  if $\alpha=2$ then $b_{\mathrm{I,II}}=1.4$, and if $\alpha=4$ then $b_{\mathrm{I,II}}=1.2$. That is to say that the region I where the system converges to a state with $\bar\rho\approx\rho^*=1/2$ shrinks with increasing $\alpha$. Strikingly, a new absorbing state $\rho=1$ (an all-C state) is appearing in the region II. This region thus have three possible steady states $\rho=0$, $\rho^*$ and $1$, and which one the system ends up in is a probabilistic event with a probability of the various outcomes that depend on  $\alpha$ and $b_0$. In particular, there is a sub region in II where the system goes almost surely to the all-C state. For example, between $3<b_0<4$ and  $\alpha=4$. This is different from the region $\mathrm{II}^*$ (in the case of no non-local links) above. Since non-local links shrinks the distance scaling in the network (from $N^{1/2}$ to $\log N$ or even shorter)~\cite{smallworld}, C and D clusters have a larger interface, which apparently is to the C cluster's advantage. With increasing $b_0$, the probability that the system ends in all-C decreases, and vanishes completely at $b_{\mathrm{II,III}}$. In Figs.~\ref{onstatic}(d) and (f) we display typical trajectories of $\rho$ and $b$ for $b_0 =1.1$, $3.5$ and $8.0$ with $\alpha=4$. These curves indicate that the system stabilizes to a steady cooperation level after about $10$ timesteps. By the adaptive payoff dynamics, we can explain the transient oscillations. Since each node is connected to a random partner by a non-local link, it has a chance of connecting to any other node and it is not too far-fetched to assume a well-mixed approximation. If we, furthermore, assume the strategy adoption rate is proportional to the relative success of the strategies, then one can approximate the dynamics by the following replicator equation system
\begin{subequations}\label{eq:repeq}
\begin{eqnarray}
\frac{\mathrm{d}\rho}{\mathrm{d}t} &=&\left\{\begin{array}{ll}\rho^2(1-\rho)(1-b) &\mbox{if~} \rho\in[0,1]\\0 & \mbox{otherwise}\end{array}\right.\label{eq:rho_rep_eq}\\
\frac{\mathrm{d}b}{\mathrm{d}t} &=&\alpha(\rho-\rho^*).\label{eq:b}
\end{eqnarray}
\end{subequations}
From the factors $\rho^2$ and $1-\rho$ in Eq.~(\ref{eq:rho_rep_eq}), we see the fixed points $0$ and $1$ of $\rho$. The $\rho\simeq\rho^*$ fixed point, however, cannot be explained by these equations. From Eqs.~(\ref{eq:repeq}), we can understand the oscillatory behavior at least qualitatively. If we have $b>1$ and $\rho>\rho^*$, then $b$ will increase and $\rho$ decrease. After some time, this situation will make $\rho$ lower than $\rho^*$, i.e.\ $\mathrm{d}b/\mathrm{d}t$ is negative. From this situation, with a decreasing $b$ and $\rho$, we will eventually reach a situation where $b$ is less than $1$. If this happens before $\rho$ hits zero, but after it falls below $\rho^*$, then both $\rho$ and $b$ will start growing again. Overall, this describes a cyclic phenomenon, which then, in practice, dampens out to an intermediate, non-trivial fixed point, or hits the all-C or all-D fixed point. We never see any sustained oscillations. Our Java applet of the adaptive game with non-local links gives a good illustration of how these oscillations look at a configuration level~\cite{java}. Perhaps the most interesting observation in this simulation is that there is an all-cooperator state for some parameter values. As an illustration, consider the $b_0=3.5$ curve in Fig.~\ref{onstatic}(d). At $t=1$, $\bar\rho$ decreases to almost $0$. This can be explained by the strong initial temptation to defect. A few cooperators survive and seed an increasing cooperator cluster (cf.\ the replicator dynamics above).  When $\bar\rho$ becomes larger than $\rho^*$, $b$ starts increasing again, but here $b$ is still too small to stop the system from getting  absorbed by the all-C state. For large $b_0$ ($\geq 3.5$), we see that $\bar\rho$ approaches its final value monotonically. For smaller values, however, we observe the above-explained oscillations. In the interval $b_0>b_{\mathrm{I,II}}$, we see that the system oscillates to wildly for the system to respond. This has the consequence that it hits one of the fixed points. Figure~\ref{onstatic}(h) shows a diagram over the different dynamic regions of $\alpha$--$b_0$ parameter space for this case with long-range links. The inclusion of the long-range links obviously changes the region-diagram quite considerably. We note that the boundary value, $b_{\mathrm{I,II}}$, separating region I from II decreases with an increasing $\alpha$ ($b_{\mathrm{I,II}}\approx 2$ for $\alpha\leq 0.1$ and $b_{\mathrm{I,II}}\rightarrow 1$ as $\alpha$ grows towards $5$), and $b_{\mathrm{II,III}}$, separating region II from III, increases monotonically with $\alpha$. In Fig.~\ref{onstatic}(h), we see that region II becomes wider with increasing $\alpha$, meaning that in our model, the presence of non-local interactions enhances cooperation.

\begin{figure}
\includegraphics[width=\linewidth]{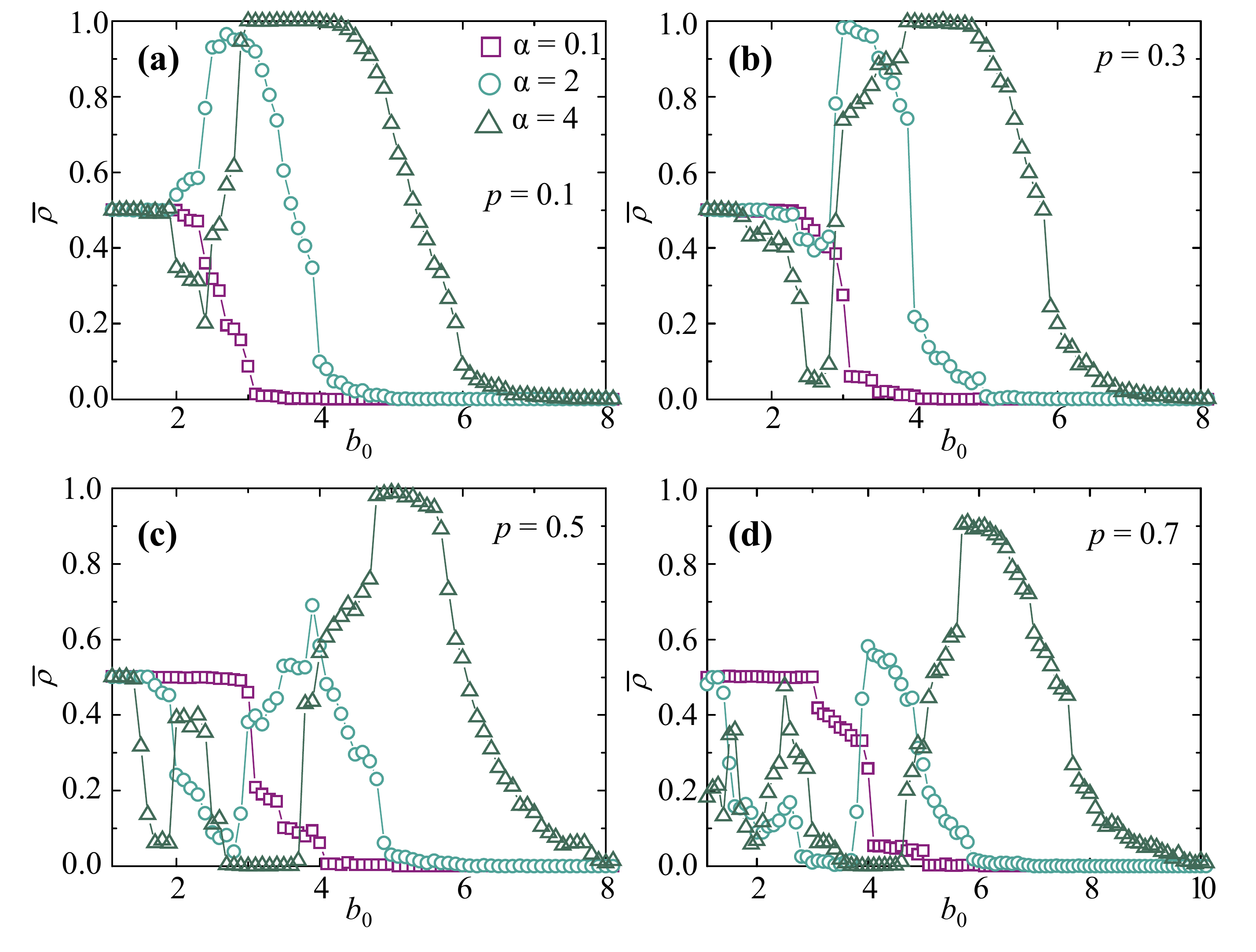}
\caption{(Color online) The average density of cooperators $\bar\rho$ of the multiadaptive model on randomly but locally connected lattice. Local link are removed with a probability $p=0.1$ (a), 0.3 (b), 0.5 (c), and 0.7 (d). We use parameters $\alpha=0.1, 2, 4$ in all panels.}\label{ranlattice}
\end{figure}

\section{Multiadaptive model}\label{mmodel}

In this section, we go one step further by considering not only how interaction determines the evolution of cooperation, but also how the interaction patterns themselves emerge, arriving at the full model of Ref.~\cite{ourPRL}. To this end, we extend our adaptive model by including a mechanism where the agents are allowed to adjust their non-local links to maximize their payoffs. As mentioned in the Introduction, we model strong ties by keeping the local interactions fixed~\cite{ourPRL}.

We update the state of the system, both strategies and non-local links, synchronously. At a timestep, each agent plays the NM game with all its local and non-local neighbors. After all agents $i$ collected their payoff, they look through their neighborhoods and, in another agent $j$ has a higher payoff, $i$ adopts $j$'s  strategy, C or D, with a probability given by Eq.~(\ref{rewiringprob}), and, if it updates the strategy, it simultaneously rewires its non-local link to the non-local neighbor of $j$. We require the graph to be simple, so we do not allow self-links and multiple links.

\begin{figure}
\includegraphics[width=\linewidth]{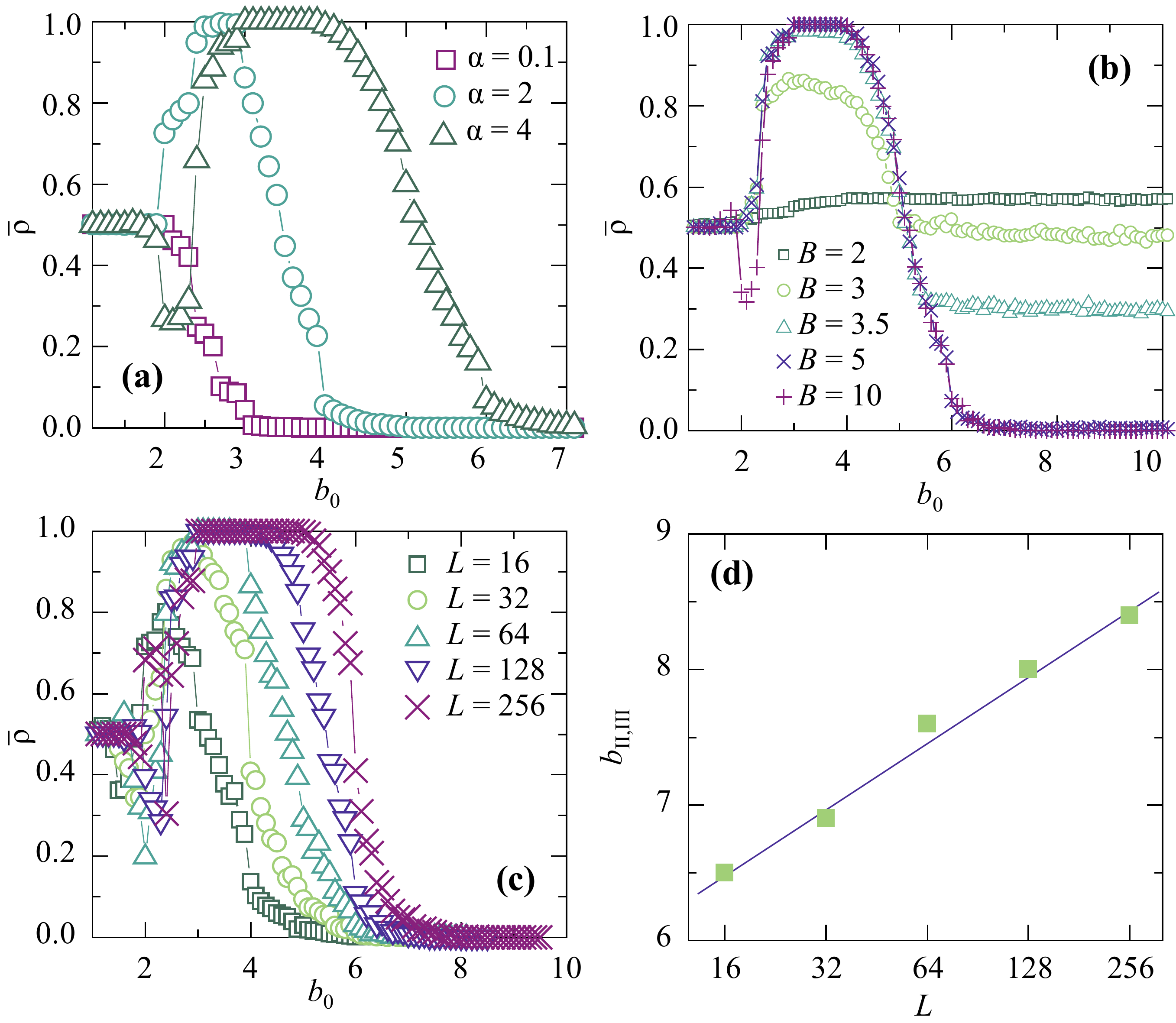}
\caption{(Color online) Panel (a) shows average density of cooperators $\bar\rho $, as a function of the initial temptation $b_0$, with $\alpha=0.1$, $2$, and $4$. Panel (b) shows effects of a bound on the temptation values as $b(t+1)=B$ for $b(t+1)>B$ and $b(t+1)=-B$ for $b(t+1)<-B$, otherwise $b(t+1)$ follows Eq.~(\ref{tupdate}). Used parameters are $\alpha=4$ and $\beta=1$. In panel (c) $\bar \rho$ is plotted as a function of $b_0$ for various size $L$ with $\alpha=4$ and $\beta=1$, and panel (d) shows that $b_{\mathrm{II,III}}$ increases logarithmically with $L$.}\label{rhocall}
\end{figure}

In Fig.~\ref{rhocall}(a), we plot the average density of cooperators $\bar\rho$ as a function of $b_0$ for three values of $\alpha=0.1$, $2$ and $4$. One can note that $\bar\rho$ shows qualitatively similar behavior to the adaptive model with non-local links discussed in Section~\ref{static}. To be more specific, also in the multiadaptive case there are  three regions, whose boundary values $b_{\mathrm{I,II}}$ ($b_{\mathrm{II,III}}$) decrease  as $\alpha$ increases. In this case too, the system reaches all-C state for certain values of $b_0$ in region II, and the time evolution of $b$ and $\rho$ shows similar transient oscillation behavior. Nonetheless, there are quantitative differences. For example, the systems with $b_0\lesssim 2$ still shows steady-state behavior ($\bar\rho\approx\rho^*=1/2$) for $\alpha=4$. Thus, in region I, for all measured values of $\alpha$, the system relaxes to a steady state with $\bar\rho\approx\rho^*$ and $\bar b$ converges to a stable value. For example, if $b_0=1.3$ we have $b(t\rightarrow\infty)\simeq 2.6$. In addition, cooperation is more strongly promoted by the adaptive networks as the defection is effectively more strongly inhibited by the feedback from the environment to the payoff matrices. 

As a generalization of multiadaptive model, we consider a probability $p$ that each local connection of two-dimensional lattice is disconnected. Thus, each agent has local links with $0\leq k_l \leq 4$ and an adjustable non-local link. If $p=0$ then it is exactly the same as multiadaptive model above. Controlling the probability $p$, we investigate multiadaptive models on the percolation cluster with non-local connections. As shown in Fig.~\ref{ranlattice}, the density of cooperators shows qualitatively similar behavior when $p$ is small. Increasing $p$, all-C state disappears when $\alpha =4$ from $p=0.5$ (two-dimensional bond percolation threshold). The boundary value separating region I from II (II from III) increases with an increasing $p$. Thus, more strong feedback is needed for the system to reach all-C state as increasing p. From these results, we find that the local connections are essential to support cooperation. 

In the following, we will investigate in detail how other factors, such as restricting  $b$, finite system size, influence the evolution of cooperation and the interaction patterns among the agents.

\subsection{Multiadaptive model with bounded temptation}

Unless the system has reached a steady-state with a finite fraction ($0<\rho<1$) of cooperators, $b$ grows (if $\rho=1$) or decreases (if $\rho=1$) unboundedly. For example, in all-C state the $b$ will increase forever according to~Eq.(\ref{eq:b}). In this situation, as the fixed points in any real system would be metastable rather than permanent, $b$ should not be overinterpreted. In principle, one can say that when the fixed point is reached is the end of applicability of the model. Alternatively, one can patch the model by imposing a bound on $b$.  We limit the temptation in Eq.~(\ref{tupdate}) by letting $b(t+1)=B$ if $b(t)>B$ and $b(t+1)=-B$ if $b(t)<-B$. The numerical results  for $\alpha=4$ of this modified model are plotted in Fig.~\ref{rhocall}(b). When $B$ is relatively large ($B=5$), we see that the average cooperator density as a function of the initial temptation looks qualitatively the same as in Fig \ref{rhocall}(a). That is, there are three different regions corresponding to the same type of dynamic behavior as in the unbounded case. In contrast, if $b$ is more restricted  ($B<5$), the region III (where the system sticks to the all-D fixed point) disappears. In addition, in  region II the probability the system hits the all-C fixed point decreases as $B$ increases. Instead, region II where $\rho$ converges to an intermediate value extends to larger $b_0$-values. This gives some perspective on the original model of Ref.~\cite{ourPRL}---if, in the unbounded multiadaptive model, the temptation to defect is too large, the system gets into the all-D absorbing state easily  because of the relatively slow regulation of $b$ compared to the faster regulation of $\rho$ [according to Eqs.~(\ref{eq:rho_rep_eq}) and (\ref{eq:b})].

\subsection{Finite-size effects}

Now we turn to a brief investigation of finite-size effects. Figure \ref{rhocall}(c) shows the results of $\bar\rho$ obtained for systems with linear size $L=16$, $64$ and $256$. The parameters are $\alpha=4$ and $\beta=1$. As shown in Figs.~\ref{rhocall}(c) and (d), for larger system size, the threshold separating the regions I and  II, $b_{\mathrm{I,II}}$, saturates with $L$, while $b_{\mathrm{II,III}}(b_0)$ increases logarithmically with $L$. In particular, $b_{\mathrm{I,II}}$ converges to 2, and $b_{\mathrm{II,III}}$ scales as $a\ln{L}$ with $a=0.70(4)$. In other  words, the cooperative regions are more stable the larger the system is.

\begin{figure*}
\includegraphics[width=0.9\linewidth]{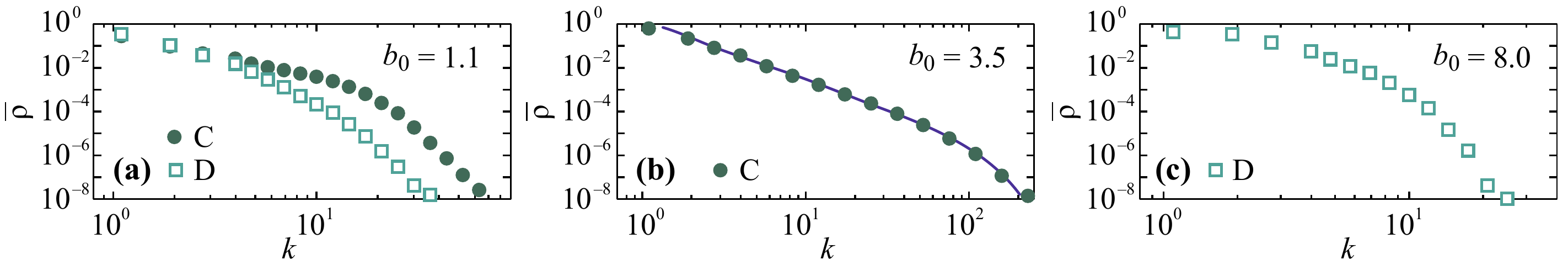}
\caption{(Color online) Correlations between the strategy and network structure. Circles (squares) correspond to the average density of cooperators (defectors) with degree $k$, $\rho_k$. Panel (a) is for $b_0=1.1$ (region I), (b) is for $b_0=3.5$ (region II), and (c) is for $b_0=8$ (region III). In panel (b) the exponent of the power-law is $2.7\pm 0.1$.}\label{rhocdk}
\end{figure*}

\subsection{Correlation between game and network structure}

In this section, we turn to the relation between game dynamics and network topology. To simplify the analysis, we will only consider the network of non-local links, disregarding the links of the background square grid. In Fig.~\ref{rhocdk}, we show $\rho_k$, the fraction of cooperators or defectors of a particular degree $k$ in the well-converged state ($t>500$). Our three different regions (as defined by the cooperator dynamics) show different network structure. For region I, represented by $b_0=1.1$ [Fig.~\ref{rhocdk}(a)], if $k\geq 3$,  cooperators have a larger $\rho_k$ than defectors. Furthermore, all nodes with $k\geq 41$ are cooperators. Since the final densities of C and D are equal in such situation, a high-degree cooperator can protect its neighbors from invasion by defectors, and thus support cooperation. For region II, e.g.\ when $b_0=3.5$ [Fig.~\ref{rhocdk}(b)] the system is stuck in the all-C state (i.e.\ $\rho_k=0$ for all $k$), we find that $\rho_k$ is fairly close to a power-law with exponential cutoff and a decay exponent is about $2.7$. Since the final state, in this case, is all-C, the payoff an agent can gather will be a linear function of its degree. Hence, during the rewiring process, the probability of getting new links of the agents will roughly be proportional to their current  degrees. In  growing networks, ``preferential attachment''---that the probability of a node to receive a new link is proportional to its degree---is known to generate a power-law degree distribution~\cite{ba:de}. In this case, where the networks are not growing Ref~\cite{salathe} shows that preferential attachment needs to be balanced by an antipreferential deletion of links in order for a power-law degree distribution to appear. In our networks, the power-law-like degree distribution remains for larger $b_0$ even though  $\rho_k$, in the steady state, varies. For systems in the all-D state, the rewiring process works differently than when $\rho=1$. Since the payoff of a defector is degree independent, then its non-local link will be rewired randomly to another D. This explains Fig.~\ref{rhocdk}(c) where we can see that the generated networks have a more narrow, peaked  degree distribution.

\begin{figure}
\includegraphics[width=\linewidth]{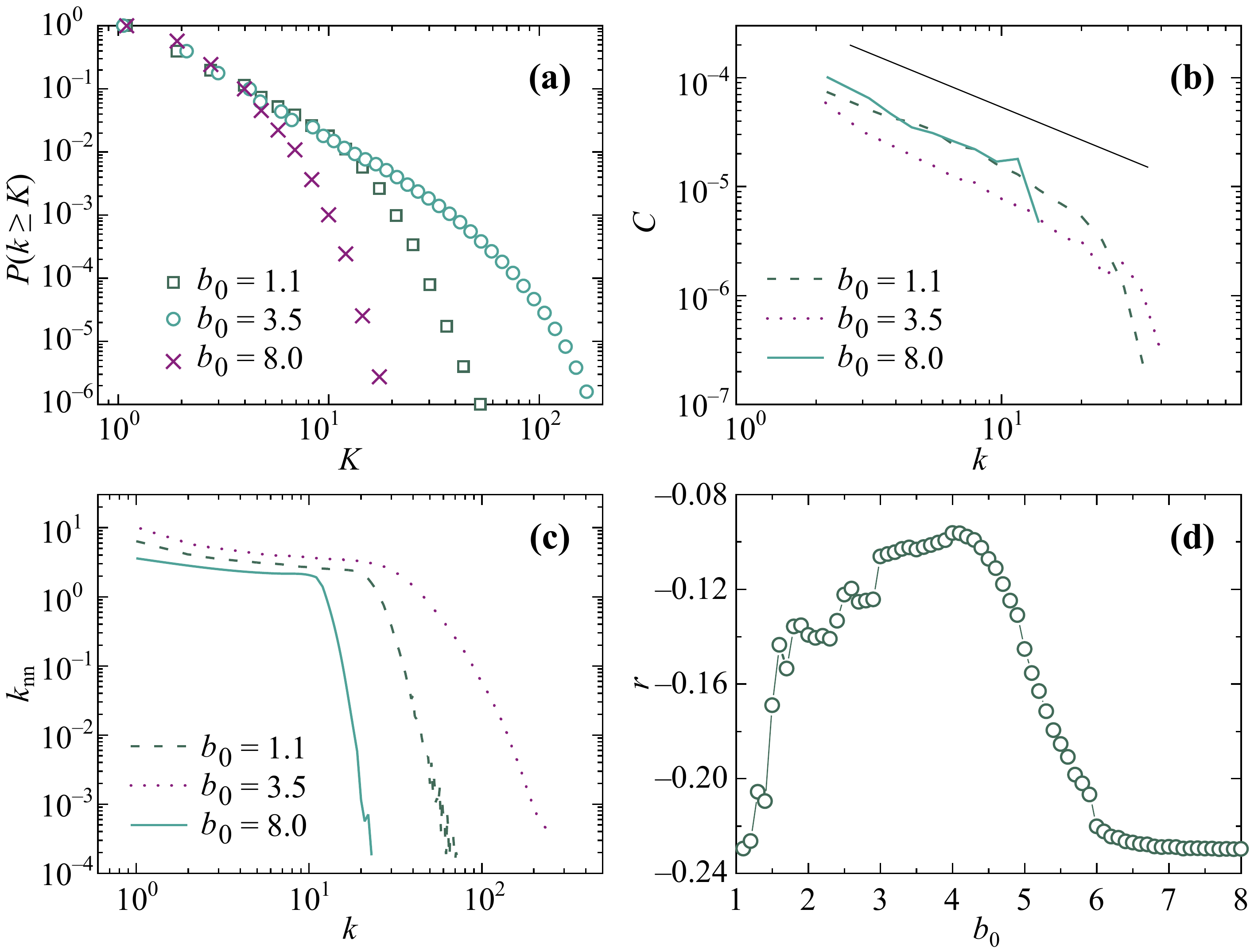}
\caption{(Color online) Structural properties of network in the steady state for different values of $b_0$ when $\alpha=4$. Panel (a) displays the cumulative degree distribution. Panel (b) shows the clustering coefficient $C$ as a function of degree $k$. The line marks a scaling inversely proportional to the degree. We investigate the degree-degree correlations by plotting the average neighbor degree $k_\mathrm{nn}$ as a function of degree $k$ (c) and the average assortativity $r$ as a function of $b_0$ (d). In plots (a)--(c) the initial temptation is $1.1$, $3.5$ and $8$ respectively. }\label{netstruct}
\end{figure}

\subsection{Emergent network structure}

Now we will turn to the network structure of the multiadaptive game model extending the analysis in Ref.~\cite{ourPRL}. From Figure~\ref{rhocdk}, we understand that  the coevolution of the contact patterns and the payoff matrix, in region II, changes the underlying network  from its initially random graph with a Poisson degree distribution to a skewed and fat-tailed degree distribution. In Fig.~\ref{netstruct}(a), we see that the cumulative degree distribution  (the probability an observed degree $k$ is larger than $K$) depends strongly on $b_0$. This is especially true for $b_0=3.5$, where the distribution seems to follow a power-law scaling for over two decades (which is quite much considering the relatively small $L=100$ system sizes). In Fig.~\ref{netstruct}(b), we make a more detailed investigation of the  hierarchical features of the steady-state networks in greater detail. This hierarchy can be characterized by the clustering coefficient (the fraction of possible triangles a node is member of with given the degree) of a node with degree $k$. If the clustering decays with degree as $C(k)\sim k^{-1}$~\cite{bara:modhie}, the network is claimed to have a hierarchical structure wherein the nodes of highest degree connected to a level below, which in turn connected to a level below, and so on. This scaling quantifies the coexistence of a hierarchical structure of nodes with different degrees. This is indeed what we observe for large $b_0$-values. To investigate the degree correlation between  nodes at either side of an link, we first measure the average degree of the nearest neighbors $k_\mathrm{nn}$ as a function of degree, $k$~\cite{Pastor2001}. If there would be no degree correlations then, $k_\mathrm{nn}(k)$ would be degree independent. This is not the case for large $b_0$-values which is disassortatively mixed, i.e.\ highly connected nodes have a tendency to be connected to low-degree nodes and vice versa. [See Fig.~\ref{netstruct}(c)]. We also measure the assortativity $r$~\cite{mejn:assmix} as a function of $b_0$ in Fig.~\ref{netstruct}(d). The assortativity confirms the conclusion from the cooperation level plots of Fig.~\ref{rhocall}---for the complex intermediate region II, the $r$ is larger than in the other regions meaning that relatively many large-degree nodes are connected to other large-degree nodes, and low-degree nodes to low-degree nodes. Metaphorically, one can see the diversity of behaviors in this region as the result from a power struggle between hubs, where the cooperator hubs win for some $b_0$-values and the defector hubs win for others.

\section{Multiadaptive model with noise}\label{wnoise}

In this section, we test extensions of  our multiadaptive model to incorporate various types of noise.

\subsection{Tuning the strategy-selection noise via $\beta$}

\begin{figure}
\includegraphics[width=0.6 \linewidth]{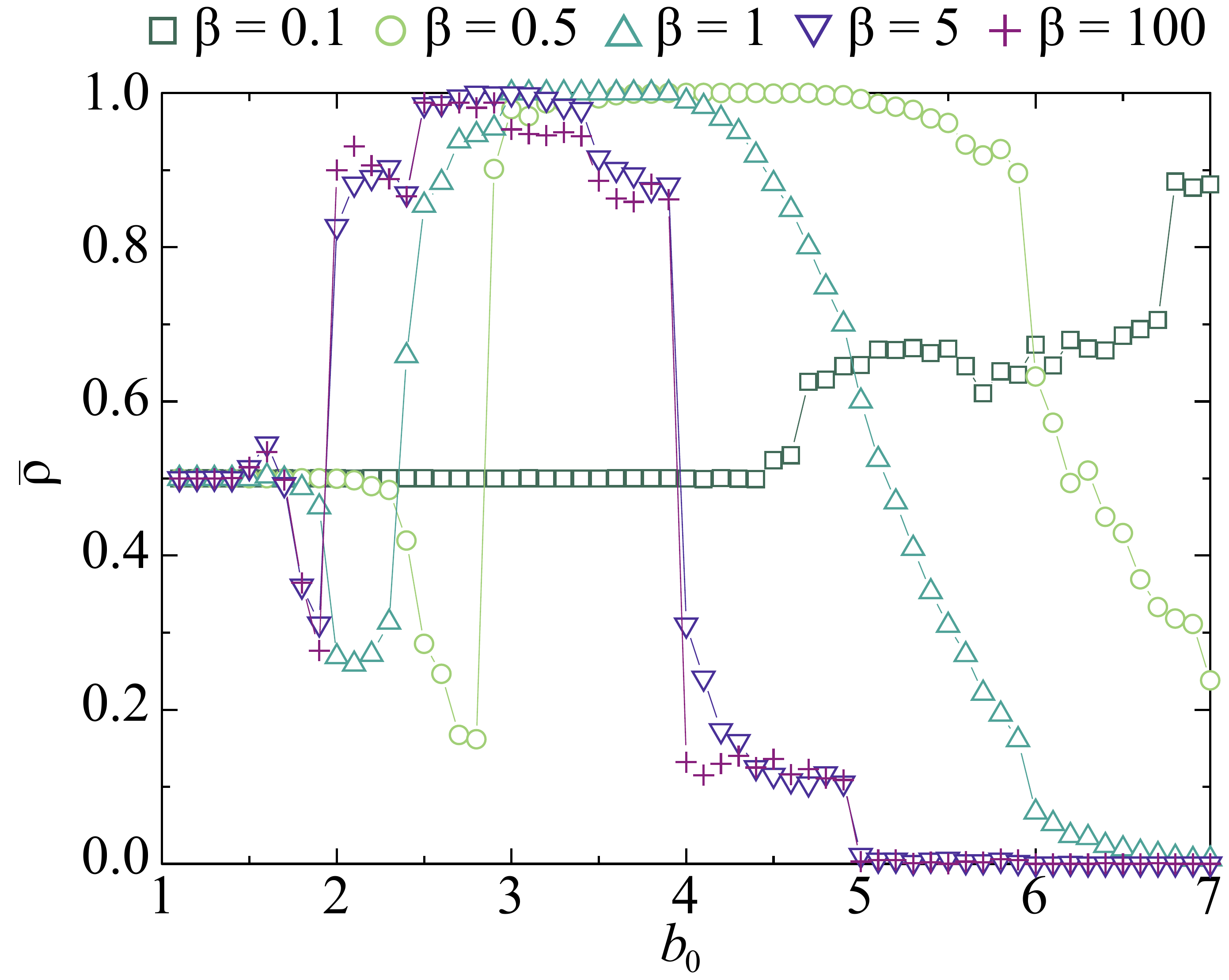}
\caption{(Color online) Response to noise in strategy selection. $\bar \rho$ as a function of $b_0$ for $\alpha=4$ with varying $\beta$.}\label{diffbeta}
\end{figure}

\begin{figure}
\includegraphics[width=0.6 \linewidth]{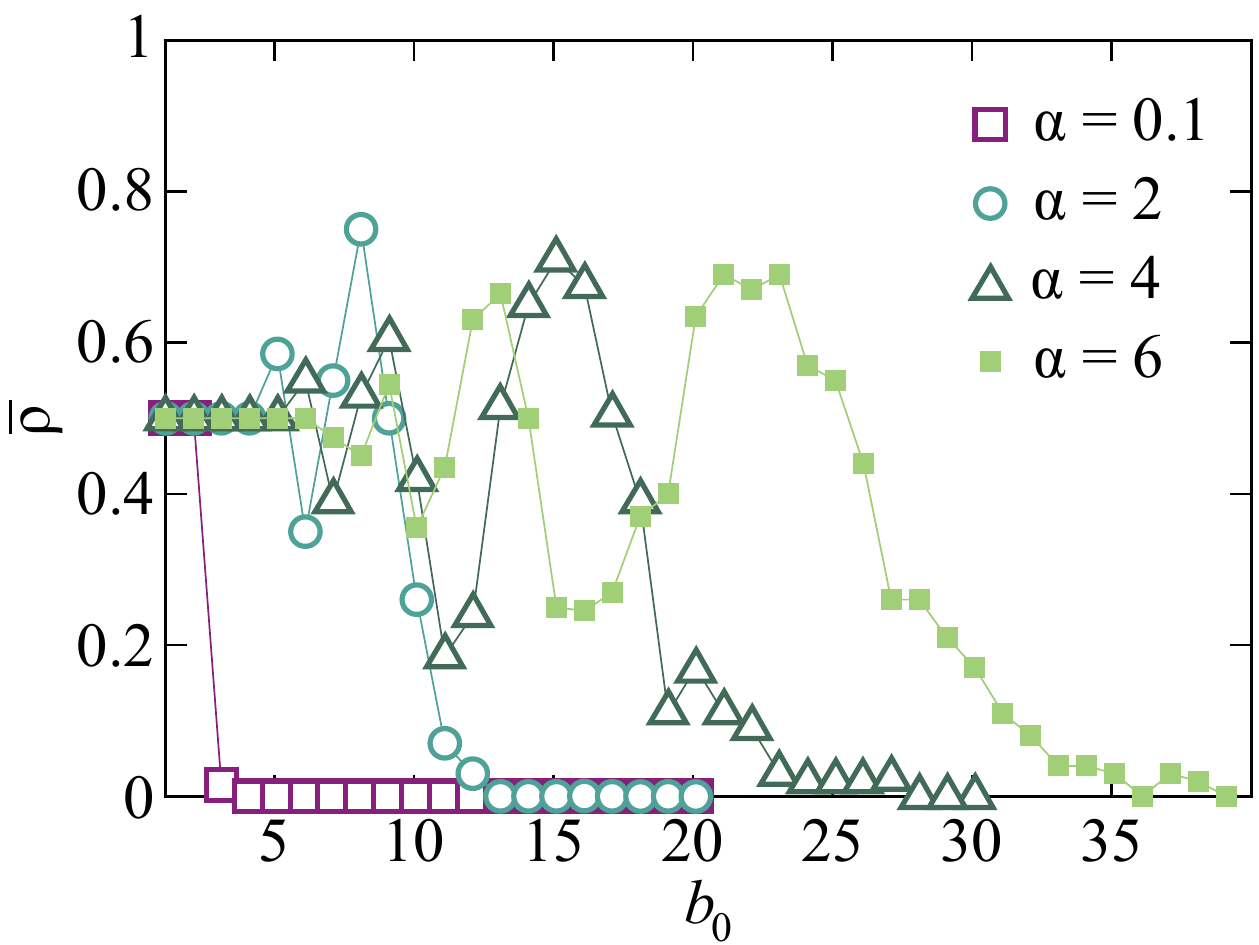}
\caption{(Color online) Noise effects on pairwise exponential comparison dynamics. We show $\bar \rho$ as a function of $b_0$ for the same parameter values as in Fig.~\ref{rhocdk}(a).}\label{pairwiseud}
\end{figure}

In our multiadaptive model, $\alpha$ and $b_0$ are the most fundamental parameters regulating the individual behavior. The parameter $\beta$ serves as a control parameter for the noise (or uncertainty) in the selection process. More concretely, it can be regarded as the reciprocal of noise intensity~\cite{szabotoke}. The larger the value of $\beta$, the less obvious the noise effect. Under the update rule of Eq.~(\ref{rewiringprob}), the strategy of a better performing neighbor is likely to be adopted, while it is also possible (if unlikely) that the strategy of a worse performing neighbor is preferred occasionally. In the limit of $\beta\rightarrow 0$ all information is lost, that is, the agents are unable to retrieve useful information from the interaction, and just switch to their strategies as by tossing a coin. In order to study the noise effect on the evolution of cooperation, we calculate how $\bar\rho$ changes by varying $\beta$. Figure~\ref{diffbeta} shows $\bar\rho$ as a function of $b_0$ for different $\beta$ in the case of $\alpha=4$. Qualitatively, $\bar\rho$ versus different $b_0$ are similar, we find visible quantitative difference among them. In particular, the range of the region II where the system evolving to all-C state expands to large $b_0$ regime as $\beta$ decreases, and the threshold value $b_{\mathrm{I,II}}$ separating the region I and II also increases monotonously with decreasing $\beta$.

If one replaces the hard selection criteria that we use (following the Nowak--May game~\cite{nowakmay} and much of the subsequent literature)---that an agents can only imitate a neighbor that performs better than itself---by a softer probabilistic rule, following Eq.~\ref{rewiringprob} also for negative $u_j-u_i$, then the system will be more strongly affected by the noise. Potentially, such rules can break cooperative states. In Fig.~\ref{pairwiseud}, we update the simulation by such a random pairwise comparison dynamics and notice that the all-C state of Fig.~\ref{rhocdk}(a) is replaced by regions of alternating higher and lower $\bar\rho$. Still, in some of these regions the cooperation level is well over 50\% and independent of $\alpha$ above some threshold (just like Fig.~\ref{pairwiseud}). We leave it for future studies to investigate the origin of the complex $b_0$ dependence in the intermediate region and whether or not the model can reach the all-C state with this type of updating dynamics.

\subsection{Strategy mutation}

In order to investigate the response of the all-C state to noise, we proceed by adding a stochastic change in the strategies. For simplicity, we start from a system consisting of only cooperators in the steady state by setting the parameters $\alpha=4$, $\beta=1$, and $b_0=3.5$. We try two cases where we flip the strategy from D to C or C to D once every hundredth timestep,  either  at the node of largest degree, or a random node. We also test a case where the nodes flip with a random change. The reason for the mutation rate is that we want to make the system able to recuperate to an all-C state (which takes less but about 100 timesteps~\cite{ourPRL}).

\begin{figure}
\includegraphics[width=0.7\linewidth]{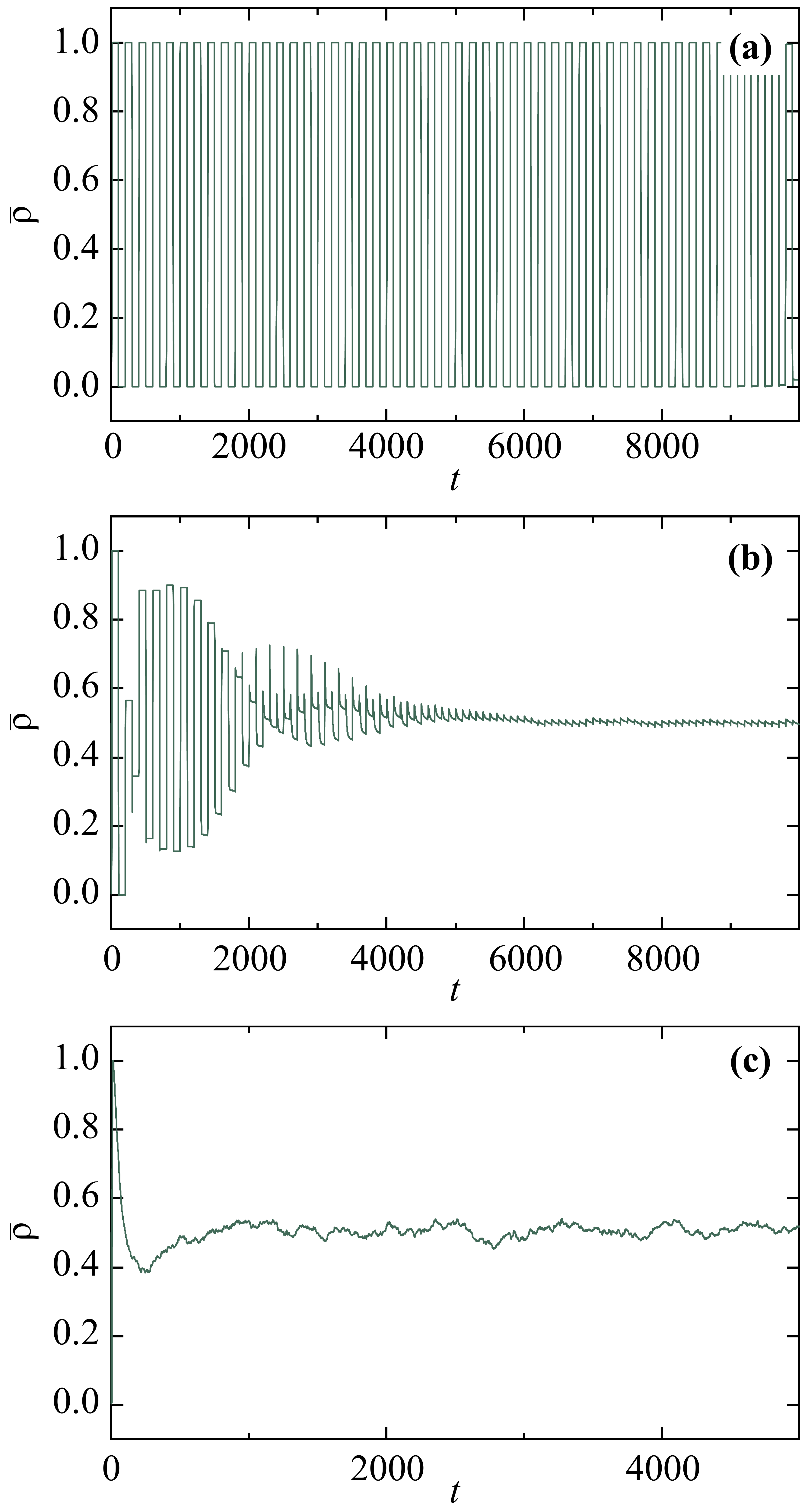}
\caption{(Color online) Three cases of strategy noise via mutation (change from D to C or C to D). The time evolution of $\bar \rho$ for different types of perturbations under the parameterizations $\alpha=4$, $\beta=1$, and $b_0=3.5$. The strategy of an agent, who is placed on the largest hub (a) or randomly selected node (b), is changed to the opposite (flipping) for every time interval $\varDelta t=100$. In the panel (c), each agent has a probability $2\times 10^{-6}$ per timestep to  regardless of payoffs.}\label{noise}
\end{figure}

In the first case, the agent located on the node with the largest degree changes its strategy to the opposite for every time interval $\varDelta$. Here, we set $\varDelta$ to $100$, which we believe is reasonable since the cooperation density of the multiadaptive model is stabilized after about 50 timesteps of relaxation. As shown in Fig.~\ref{noise}(a), if D appears on the largest hub will be rapidly spread to the whole system, and as a result all-C state changes to all-D state. On the other hand, we observe similar phenomenon if a cooperator appears on the largest-degree hub in an environment of all-D members---all D state will be change to all-C sate shortly after the perturbation. Thus, the system is alternatively switching between all-C and all-D.

In the second case, we apply the above perturbation to a randomly selected node. From Fig.~\ref{rhocdk}(b), we know that the final interaction network has a degree distribution similar to a power-law. This means that most of the players have small neighborhood. Consequently, the disturbance is most likely to affect nodes of low degree. In such situation, a disturbance through mutation (or mistakes) cannot spread to the whole system. This is because  the high-degree cooperators can protect their neighbors from imitating defectors, like the behavior seen in the region I of the multiadaptive model without noise, where $\bar\rho$ approaches to $\rho^*$ as $t$ increases [See Fig.~\ref{noise}(b)].

The last type of perturbation we consider is that each agent has a probability $2\times 10^{-6}$ to mutate per timestep regardless of payoffs. Figure~\ref{noise}(c) shows the time evolution of $\bar\rho$ when applying this type of perturbation to an all-C state. A typical picture can be seen in Fig.~\ref{noise}(b), the high degree C agents and their neighbors are not affected so much since the strategy of agents placed on nodes with low degree are mainly mutated in the most of time, and $\bar\rho$ fluctuates around $\rho^*$ as $t$ increases. Taken together, the all-C state needs to be stabilized by hubs; once the hubs change their strategies by mutation, the all-C state is no longer stable.

\section{Time scales}\label{tscale}

Up to now, we only investigated the case where the dynamical variables in our multiadaptive model evolve with the same timescale for network updates, strategy updates and feedback to from the payoff of agents to the payoff matrix ($b$, to be specific). This similarity of timescales makes sense especially in the context of evolution and population biology where one timestep of the simulation corresponds to one generation (so the time is naturally discrete).  It could be appropriate in social systems too, whenever processes happen at the similar timescales. However, there are also socioeconomic situations that would be better modeled as having different timescales for different processes. A natural way of implementing this is to use asynchronous updating where, at every timestep, one randomly chosen agent changes its strategy. By this method, one can easily tune the timescales of the processes. Another option is to use partially synchronous updates, where one of the steps is synchronous, others asynchronous.

\begin{figure}
\includegraphics[width=0.7\linewidth]{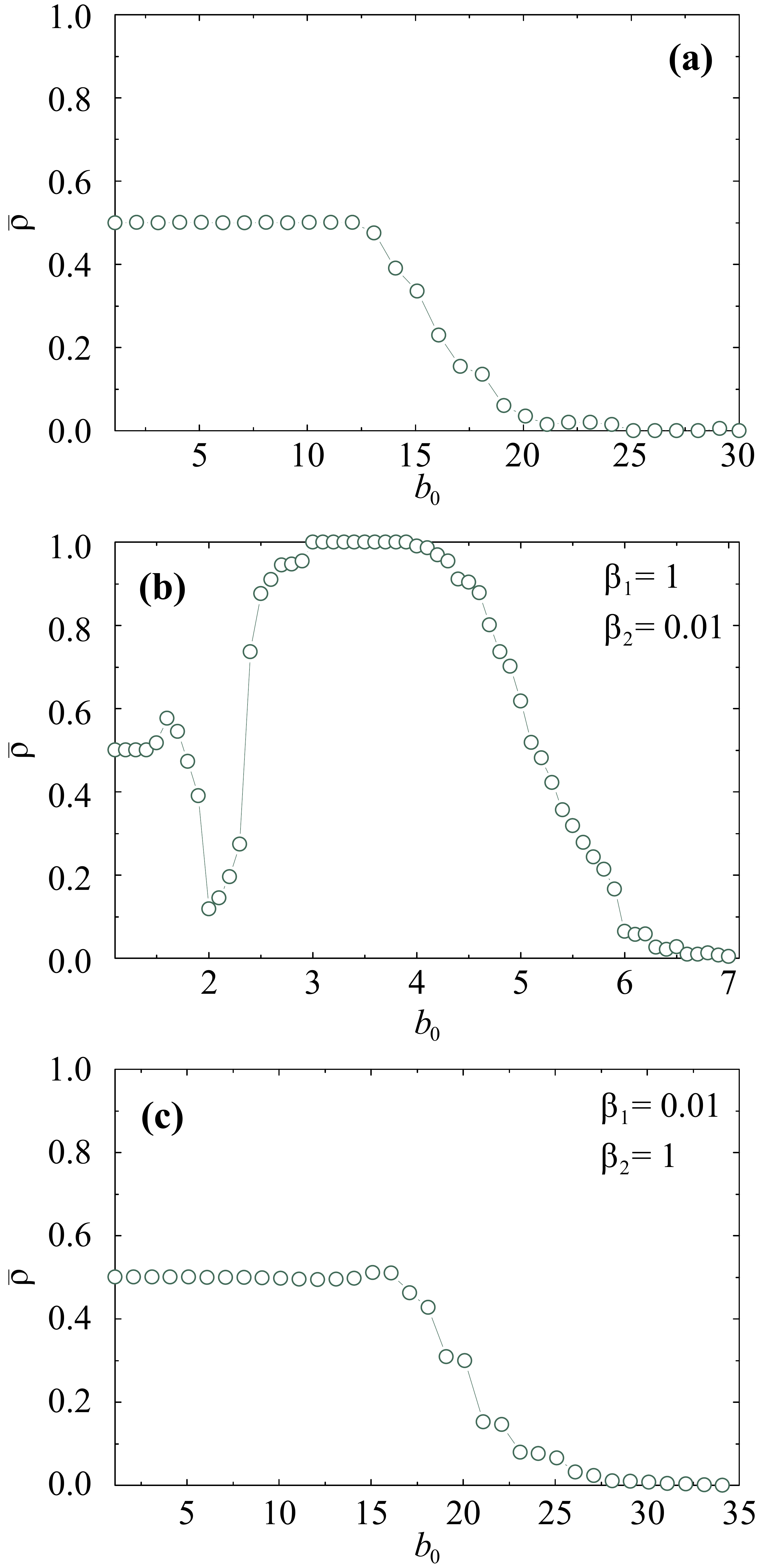}
\caption{(Color online.) An investigation of the effect of relative timescales. The average density of cooperators for random updating (a), more frequent strategy updating ($\beta_1=1$, $\beta_2=0.01$) (b), and more frequent link rewiring ($\beta_1=0.01$, $\beta_2=1$) (c). We set the parameter $\alpha$ to $4$.}\label{timescale}
\end{figure}

In this section, we consider distinct timescales for the updating of agents' strategies, the evolution of temptation and the interaction structures. We first study the case of random asynchronous updating. Here we go through all agents in a random order (the order is different from timestep to timestep). For each agent we  updating its strategy and immediately thereafter we readjust $b$ and perform the rewiring of the non-local link. We find that under the random asynchronous updating scheme,  the all-C state disappears from region II. This is in contrast to that in the case of synchronous updating; see Fig.~\ref{timescale}(a). This effect of the random update is probably due to that now  agents have more information, on average, to guide their  decisions which strategy to use. Next, we tune the relative timescales of the network and strategy updating. To do this, we separate the  probabilities for updating strategies and the social network, let $P_1(i\rightarrow j)$ represent strategy updating and $P_2(i\rightarrow j)$ for links rewiring, where
\begin{equation}
P_{1,2}(i \rightarrow j)=\frac{1}{1+e^{-\beta_{1,2}(u_j-u_i)}} .\label{separateprob}
\end{equation}
 The parameters $\beta_1$ and $\beta_2$ control the probabilities (hence the speed, or timescale) of the evolution. Let us define the average time $\tau_1$  for link rewiring to be occurred once. The average number of link rewiring occurred until time $t$, $n_1$, equals to $tP_1$. The average time $\tau_1$ is in inverse proportional to $n_1$, and then $\tau_1 \sim \frac{1}{n_1}\sim \frac{1}{P_1}$. Using Eq. (\ref{separateprob}), we get $\tau_1 \sim 1+\exp^{-\beta_1 \Delta u}$, where $\Delta u =u_j - u_i$. Because we use best rule, the neighborÕs payoff is always larger than $i$Õs, $u_j \geq u_i$. In sum, the average time for link rewiring occurred once is a decreasing function of  $\beta_1$.  By the same calculation, the average time for strategy updating occurred once is decreasing function of $\beta_2$. Thus, we can control the time scale of strategy updating and link rewiring separately with $\beta_1$ and $\beta_2$. Given the feedback strength $\alpha$, the system has all-C state if the inequality $\beta_1 \geq \beta_2$ is satisfied. Figures~\ref{timescale}(b) and (c) show $\bar\rho$ as a function of $b_0$ for two different sets of parameters $(\beta_1,\beta_2)=(1,0.01)$ and $(0.01,1)$ respectively. When strategy updating is more frequent than link rewiring (i.e.\ $\beta_1>\beta_2$), we observe a similar behavior of $\bar\rho$ versus $b_0$ as  in Fig.~\ref{rhocall}(a). On the other hand, when link updating is more frequent than strategy updating, the system never reaches the all-C boundary; see Fig.~\ref{timescale}(c). In  region II, the system reaches either the $\rho=\rho^*$ or the $\rho=0$ state, and the probability getting into all-D state increases with increasing $b_0$. The effect of more frequent link updating is similar to the  random dynamics in the sense that the all-C state is lost. This  suggests that the random dynamics efficiently slows down the updating of strategies. From this, we learn that the existence of the all-C state requires a comparatively faster strategy dynamics compared to the link dynamics.

\section{Discussion}\label{discussion}

In this paper, we have studied a game-theoretical model with feedback from the behavior of the agents to the rules of the game, via the payoff matrix, and an active optimization of both the contact structure between the agents and their strategies. We investigate this model, first presented in Ref.~\cite{ourPRL}, by extending it in many ways. With respect to the average cooperation density, the model is a nonequilibrium model (in the statistical-mechanics sense). This makes the initial temptation value $b_0$ a crucial model parameter. Like in Ref.~\cite{ourPRL}, we  identify three regions of distinct dynamic behavior for a large parameter space and different generalizations of the model. In region I, the average cooperator density relaxes to a stable level through damped oscillations; in region III the systems reaches an all-defect state. For intermediate $b_0$-values (region II), the system ends at one of three fixed points, $0$, $\rho^*$ or $1$, with parameter-dependent probabilities. For some parameter values in this region, the original multiadaptive model  will almost certainly reach an all-cooperator state. This all-cooperator state is absorbing, but when we extend the model by adding noise, this state rarely appears. More precisely, if the hubs of the network can mutate their strategies, the all-C state will not be stable---the all-C state needs to be stabilized by cooperator hubs. When we tune the timescales between link and strategy updates, we find that the all-C state needs a faster strategy update; if the link dynamics is to frequent, then the all-C state is instable. An interesting aspect of the all-C state is that has power-law like degree distribution with a  $C\sim 1/k$ scaling of the clustering coefficient (a hallmark of hierarchical organization~\cite{bara:modhie}). Traditionally, hierarchies are usually explained as consequences of factors external to the social system, e.g.\ age or fitness~\cite{wilson}. 

We use several different updating rules---random updating with following the best (Fig.~7(a)) and updating rule with mutation probability (Fig.~6(c)). Additionally, we investigate the model with different updating rule, in which each agent chooses the random neighbor and imitates his strategy with the probability of Eq. \ref{rewiringprob}. The system doesnÕt go to all-C state anymore (not shown). We think this result is related to the time scale of link rewiring. Under this updating rule, an agent needs more time to find the most profitable neighbor with non-local link. This means that link rewiring is effectively slower than strategy updating. However, since this time-scale difference isnÕt explicit the average cooperation has high level (not all-C) in some range of $b_0$.  

In the case of $\alpha < 0$, we can expect the result obviously. There is only one directional feedback, accelerating cooperation or defection. Since the temptation of defect at initial time ($t=0$) is larger than 1 and the temptation is increasing for $\rho_C < \rho^*$, agents prefer to act as defector and $b$ is always increasing. Finally, the system always goes to all-D state. On the other hand, we can think quite narrow region in $\alpha<0$ case. Suppose that $b_0$ is small (for example, $b_0 =1.01$), $\rho_C(0) \gg \rho_D(0)$ (for example, $\rho_C(0)=0.99$), and negatively strong alpha (ex. $\alpha= -4$). In this setting, the system reaches the all-C state. 

To epitomize, our work shows a generalization of spatial social dilemma models where hierarchies can emerge in a cooperative state. In our framework, these hierarchies need stable cooperating hubs to persist. In this sense, the hierarchies are more the result of an all-cooperative state than a prerequisite for its emergence. We note that in the literature, there are conflicting results on whether or not hierarchy promotes cooperation, or not~\cite{vukov:pd_santos:pd}---in different games the effect can be different.

 We believe there are many interesting multiadaptive directions to for Nowak--May type spatial games. In this work  the interaction network and the payoff matrix is controlled by the game, one can also imagine situations when the dynamics (the timescales) is a outcome of the game dynamics and the agents are more heterogeneous (so their payoff can be reinvested into their ability to play the game).

\acknowledgments{This research was supported by the Wenner--Gren Foundations (SL), the Swedish Research Council (PH), and the WCU program through NRF Korea funded by MEST R31--2008--10029 (PH), and the National Natural Science Foundation of China, Grant Nos.\ 11005051 and 11047606 (ZXW).}

\appendix*

\section{Algorithmic description}
In this appendix we give a more detailed description of the multiadaptive model of Section~\ref{mmodel}.
The initial state of the model is generated as follows.
\begin{enumerate}
\item Construct a $L\times L$ square lattice with closed boundary conditions (so the corner vertices interact with two others and the edge vertices interact with three others).
\item Connect every vertex with a random other vertex.
\item Assign strategies C or D randomly to all vertices.
\item Set $b(t=0)=b_0$.
\end{enumerate}
Then, from the initial state the system is updated by doing what is listed below. We describe the transition from timestep $t$ to $t+1$. 
\begin{enumerate}
\item For every vertex (in arbitrary order), calculate the payoff with the interaction with the neighbors by the Nowak--May rulesÑan interaction between a D and C contributes with 1 to the score of the defecting vertex and 0 to the cooperator, an interaction between two C gives $b(t)$ to both while two $D$ gives no profit.
\item Go through all vertices and let them copy the strategy of the neighbor (including themselves) that has the largest payoff in the previous step, also rewire the long-range link to the neighbor (excluding themselves) that has the largest payoff.
\item Calculate $b(t+1)$ by Eq.~\ref{tupdate}.
\end{enumerate}

\end{document}